\documentclass[aps,pre,showpacs,groupedaddress,twocolumn]{revtex4}

\usepackage{amsfonts}
\usepackage{mathrsfs}
\usepackage{graphicx}
\usepackage{amsmath}
\usepackage{amssymb}
\usepackage{amsbsy}

\begin{document}

\title{State sampling dependence of the Hopfield network inference}

\author{Haiping Huang$^{1,2}$}
\email{hphuang@itp.ac.cn} \affiliation{$^1$Key Laboratory of
Frontiers in Theoretical Physics, Institute of Theoretical Physics,
Chinese Academy of Sciences, Beijing 100190, China\\
$^2$Department
of Physics, The Hong Kong University of Science and Technology, Hong
Kong, China}
\date{\today}

\begin{abstract}
The fully connected Hopfield network is inferred based on observed
magnetizations and pairwise correlations. We present the system in
the glassy phase with low temperature and high memory load. We find
that the inference error is very sensitive to the form of state
sampling. When a single state is sampled to compute magnetizations
and correlations, the inference error is almost indistinguishable
irrespective of the sampled state. However, the error can be greatly
reduced if the data is collected with state transitions. Our result
holds for different disorder samples and accounts for the previously
observed large fluctuations of inference error at low temperatures.

\end{abstract}


 \maketitle

\section{Introduction}
The inverse Ising problem, also known as Boltzmann machine learning,
is recently widely studied in the context of network inference. As
we know, a large number of elements interacting with each other may
yield collective behavior at the network level. Encouragingly, the
pairwise Ising model was shown to be able to capture most of
correlation structure in real neuronal
networks~\cite{Nature-06,Tang-2008}. The advent of techniques for
multi-electrode recording or microarray measurement produces high
throughput biological data. The inverse Ising problem tries to
construct a statistical mechanics description of the original system
directly from these data, which helps to better understand how the
brain or other biological networks represent and process
information~\cite{Tkacik-2010}. On the other hand, to test proposed
efficient inverse algorithms, one can alternatively collect the
required data, i.e., magnetizations $\{m_{i}\}$ and two-point
connected correlations $\{C_{ij}\}$($i,j$ run from $1$ to $N$ and
$N$ is the number of elements in the network) from Monte Carlo
simulations of a toy
model~\cite{Aurell-2010,Marinari-2010,Huang-2010a,Huang-2010b}.
Given the magnetizations and correlations, the underlying parameters
(i.e., couplings and fields) of the pairwise Ising model are
inferred to describe the statistics of the experimental data. In
other words, the data is fitted with $P_{{\rm
Ising}}(\boldsymbol{\sigma})\propto\exp\left[\sum_{i<j}J_{ij}\sigma_{i}\sigma_{j}+\sum_{i}h_{i}\sigma_{i}\right]$,
such that the predicted magnetizations and correlations are
consistent with those measured, i.e., $\left<\sigma_{i}\right>_{{\rm
Ising}}=\left<\sigma_{i}\right>_{{\rm data}},
\left<\sigma_{i}\sigma_{j}\right>_{{\rm
Ising}}=\left<\sigma_{i}\sigma_{j}\right>_{{\rm data}}$. In this
setting, we use $\boldsymbol{\sigma}$ to represent the configuration
of the system and each component takes $\pm1$. Previous studies
along this line focused on the Sherrington-Kirkpatrick (SK)
model~\cite{Aurell-2010,Marinari-2010} and the Hopfield
model~\cite{Huang-2010a,Huang-2010b}. However, the influence of
state sampling on the network inference was overlooked and in this
work, we will illustrate this most important issue on the fully
connected Hopfield network reconstruction. We find that the quality
of reconstruction depends on the way the data is collected via state
samplings. A lazy Glauber dynamics can be easily trapped by a
high-lying metastable state, however, in a finite system, it still
has the possibility of a transition to a different state (free
energy valley), provided that the amount of sampling time is chosen
appropriately~\cite{Marinari-2011}. If we present the system at low
enough temperature $T$ and high memory load $\alpha$, these two
different scenarios for state sampling will yield different
qualities of network inference. The former maintains a high
inference error regardless of which state we sample, while the
latter reduces the error substantially.

The paper proceeds as follows. The fully connected Hopfield network
is defined in Sec.~\ref{sec_FHopf}. We collect the data using a lazy
Glauber dynamics and infer the network by message passing algorithm,
which is also demonstrated in this section. Results and discussions
are given in Sec.~\ref{sec_result}. We conclude this paper in
Sec.~\ref{sec_Conc}.

\section{Fully connected Hopfield network and its inference}
\label{sec_FHopf}

The Hopfield network has been proposed in
Refs.~\cite{Hopfield-1982,Amit-198501} as an abstraction of
biological memory storage and was found to be able to store up to
$0.144N$ random unbiased patterns~\cite{Amit-1987}. If the stored
patterns are dynamically stable, then the network is able to provide
associative memory and its equilibrium behavior is described by the
following Hamiltonian:
\begin{equation}\label{Hami}
    \mathcal{H}=-\sum_{i<j}J_{ij}\sigma_{i}\sigma_{j}
\end{equation}
where $\sigma_{i}=+1$ indicates the spiking of neuron $i$ while
$\sigma_{i}=-1$ means the silence. Coupling between neuron $i$ and
$j$ is symmetric and constructed according to the Hebb's rule:
\begin{equation}\label{Hebb}
     J_{ij}=\frac{1}{N}\sum_{\mu=1}^{P}\xi_{i}^{\mu}\xi_{j}^{\mu}
\end{equation}
where $\{\xi_{i}^{\mu}\}$ are $P$ stored patterns with each element
$\xi_{i}^{\mu}$ taking $\pm 1$ with equal probability. These random
stored patterns give rise to disorder leading to frustration in the
low temperature. The ratio of the number of stored patterns to the
network size $N$ is defined as the memory load $\alpha$, i.e.,
$\alpha\equiv\frac{P}{N}$.

Our prime concern is the study of the fully connected Hopfield
network inference. In this network, each neuron is connected to all
the other neurons and no self-interactions and external fields are
assumed. The equilibrium properties of the fully connected Hopfield
model has been addressed in Ref.~\cite{Amit-1987}. In this work, we
focus on the glassy phase which takes place when
$T<T_{g}=1+\sqrt{\alpha}$. At all finite $\alpha$, this phase has a
vanishing small overlap with any of the stored patterns.
Furthermore, the replica symmetry solution for this phase is
unstable and develops a hierarchically organized
structure~\cite{Amit-1987,Toki-93} which leads to anomalously slow
dynamic relaxation. The dynamics was shown to exhibit aging
phenomena supporting the nontrivial structure of the phase
space~\cite{Cannas-2000}. Therefore, starting from different initial
configurations, the lazy Glauber dynamics will be trapped in
different free energy valleys with high probability. For a finite
system, free energy barriers around metastable states are always
finite and the Glauber dynamics has the possibility to escape from
local minima of free energy landscape. Therefore, as an inverse
problem, we are interested in the influence of state sampling on the
network inference in this phase, and we look at individual disorder
samples with $N=125,\alpha=0.2$ in the low temperature $T=0.5$, and
expect analysis on these individual disorder samples will provide
valuable information on the state sampling dependence of the network
inference for a general context.

To sample the state of the original model Eq.~(\ref{Hami}), we apply
a lazy (non-optimized) Glauber dynamics rule:
\begin{equation}\label{GDrule}
    P(\sigma_{i}\rightarrow-\sigma_{i})=\frac{1}{2}\left[1-\sigma_{i}\tanh\beta h_{i}\right]
\end{equation}
where $\beta$ is the inverse temperature and $h_{i}=\sum_{j\neq
i}J_{ij}\sigma_{j}$ is the local field neuron $i$ feels. In
practice, we first randomly generate a configuration which is then
updated by the local dynamics rule Eq.~(\ref{GDrule}) in a randomly
asynchronous fashion. In this setting, we define a Glauber dynamics
step as $N$ proposed flips. As a lazy dynamics, we quench the system
directly to the preset low temperature $T=0.5$ without any annealing
schemes and run totally $4\times 10^{6}$ steps, among which the
first $2\times 10^{6}$ steps are run for thermal equilibration and
the other $2\times 10^{6}$ steps for computing magnetizations and
correlations, i.e., ${m_{i}=\left<\sigma_{i}\right>_{{\rm data}},
C_{ij}=\left<\sigma_{i}\sigma_{j}\right>_{{\rm data}}-m_{i}m_{j}}$
where $\left<\cdots\right>_{{\rm data}}$ denotes the average over
the collected data. The state of the network is sampled every $100$
steps after thermal equilibration.

Given the measured magnetizations and correlations, we attempt to
infer couplings via susceptibility propagation (SusProp) update
rule~\cite{Mezard-09} which was shown to outperform other
mean-field-type methods~\cite{Huang-2010b}. Before introducing this
rule, we define two kinds of relevant messages. One is the cavity
magnetization $m_{i\rightarrow j}$ of neuron $i$ in absence of
neuron $j$; the other is the cavity susceptibility $g_{i\rightarrow
j,k}$ which is the response of the cavity field $h_{i\rightarrow j}$
to the small change of the local field $h_{k}$ of neuron $k$. The
SusProp rule can be derived using belief propagation plus
fluctuation-response relation~\cite{Huang-2010b} and is formulated
as follows:
\begin{subequations}\label{SusP}
\begin{align}
m_{i\rightarrow j}&=\frac{m_{i}-m_{j\rightarrow i}\tanh J_{ij}}{1-m_{i}m_{j\rightarrow i}\tanh J_{ij}}\\
g_{i\rightarrow j,k}&=\delta_{ik}+\sum_{n\in \partial i\backslash
j}\frac{1-m_{n\rightarrow i}^{2}}{1-(m_{n\rightarrow i}\tanh
J_{ni})^{2}}
\tanh J_{ni}g_{n\rightarrow i,k}\\
J_{ij}^{{\rm
new}}&=\epsilon\left[\frac{1}{2}\log\left(\frac{(1+\widetilde{C_{ij}})(1-m_{i\rightarrow
j}m_{j\rightarrow i})}
{(1-\widetilde{C_{ij}})(1+m_{i\rightarrow j}m_{j\rightarrow i})}\right)\right]+(1-\epsilon)J_{ij}^{{\rm old}}\\
\widetilde{C_{ij}}&=\frac{C_{ij}-(1-m_{i}^{2})g_{i\rightarrow
j,j}}{g_{j\rightarrow i,j}}+m_{i}m_{j}
\end{align}
\end{subequations}
where $\partial i\backslash j$ denotes neighbors of neuron $i$
except $j$, $\delta_{ik}$ is the Kronecker delta function and
$\epsilon (\in[0,1])$ is introduced as a damping factor and should
be appropriately chosen to prevent the absolute updated
$\tanh(J_{ij})$ from being larger than $1$. Note that all couplings
in Eq.~(\ref{SusP}) have been scaled by the inverse temperature
$\beta$.

To evaluate the reconstruction performance of SusProp, we define the
inference error as
$\Delta=\left[\frac{2}{N(N-1)}\sum_{i<j}(J_{ij}^{*}-J_{ij}^{{\rm
true}})^{2}\right]^{1/2}$ where $J_{ij}^{*}$ is the inferred
coupling while $J_{ij}^{{\rm true}}$ is the true one constructed
according to Eq.~(\ref{Hebb}). In Eq.~(\ref{SusP}),
$\left(\{m_{i}\},\{C_{ij}\}\right)$ serve as inputs to the update
rule. To run SusProp, we initially set all couplings to be zero and
randomly initialize for every directed edge the message
$m_{i\rightarrow j}\in [-1.0,1.0]$ and $g_{i\rightarrow j,k}=0$ if
$i\neq k$ and $1.0$ otherwise. Then SusProp is iterated according to
Eq.~(\ref{SusP}) until either all inferred couplings converge within
a preset precision $\eta$ or the maximal number of iterations
$\mathcal {T}_{max}$ is reached. In practice, we set $\mathcal
{T}_{max}=3000,\eta=10^{-4}$ and $\epsilon$ varies from
$\mathcal{O}(10^{-2})$ to $\mathcal{O}(10^{-4})$.

\section{Results and discussions}
\label{sec_result}

\begin{figure}
    \includegraphics[width=0.5\textwidth]{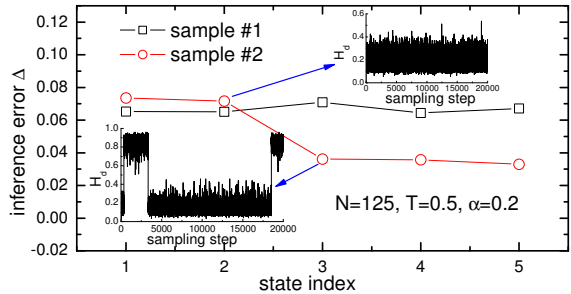}
  \caption{
  (Color online) State sampling dependence of the network inference. We present the inference error for two
  distinct disorder samples. For the second sample, the first two state samplings give large inference errors while
  the last three samplings reduce the error substantially. Insets
  give their corresponding evolutions of Hamming distances between the current sampled configuration and the
  first sampled one. For the first sample, all five state
  samplings provide nearly the same inference errors and their
  evolutions of the Hamming distances are similar to that of the
  second state sampling of the second sample. Note that the state
  index for the last three samplings of the second sample means
  different paths of multiple states sampling and that for the others
  means different states where the sampling is confined.
  }\label{state_sampling}
\end{figure}
We simulate the fully connected Hopfield network of size $N=125$ at
$T=0.5$ and $\alpha=0.2$ forcing the system to enter the glassy
phase. The state sampling dependence of the network inference is
illustrated in Fig.~\ref{state_sampling}. To discriminate two kinds
of scenarios for state sampling, we track the evolution of Hamming
distance between current sampled configuration and the first sampled
one. By Hamming distance, we mean
$H_{d}=\frac{1}{2}\left(1-\frac{1}{N}\sum_{i=1}^{N}\sigma_{i}^{t}\sigma_{i}^{0}\right)$
where $\boldsymbol{\sigma}^{t}$ is the current sampled configuration
while $\boldsymbol{\sigma}^{0}$ is the first sampled one. We also
measure the energy for each sampled configuration during the whole
state sampling process. For the lower inset of
Fig.~\ref{state_sampling}, the energy fluctuates around $-0.511$
with fluctuation of order $0.023$ while around $-0.510$ for the
upper inset with nearly the same fluctuation amplitude. It can be
seen clearly that the inference error depends strongly on the way
the state is sampled, regardless of which state the lazy dynamics
visits. In the first type, the sampling is confined in a single free
energy valley, or the same level of the family tree like structure
of the phase space~\cite{Mezard-1987,Toki-93}. This case would
probably occur since the limited amount of sampling time is not
enough for the dynamics to escape from the current valley.
Therefore, we observe one mean value of Hamming distance in the
upper inset. Unfortunately, this type produces highly magnetized
data especially at the low temperature, which gives rise to the
non-convergence of SusProp and a high inference error. It should be
emphasized that all samplings with the similar feature of Hamming
distance evolution, as the upper inset shows, exhibit nearly the
same inference error irrespective of the sample and the sampled
state. In the second type, a transition to a different free energy
valley or a higher level of phase space organization may happen due
to the finiteness of the network if the temperature is not very low.
We do observe this possibility in our simulations as the lower inset
shows. In this case, another larger mean Hamming distance appears
during the state sampling. Since each free energy valley is visited
by the Glauber dynamics with a probability proportional to its
thermodynamical weight~\cite{Zhou-2007ctp}, when state transitions
occur, the computed average of $\sigma_{i}\sigma_{j}$ or
$\sigma_{i}$ over all $2\times10^{4}$ sampled configurations amounts
to the weighted sum, i.e., $\left<\sigma_{i}\sigma_{j}\right>_{{\rm
data}}=\sum_{\gamma}w^{\gamma}\left<\sigma_{i}\sigma_{j}\right>_{\gamma},\left<\sigma_{i}\right>_{{\rm
data}}=\sum_{\gamma}w^{\gamma}\left<\sigma_{i}\right>_{\gamma}$
where only a few states are considered depending on the actual state
transitions in the sampling process and $\gamma$ is the state index
the dynamics visits and $w^{\gamma}$ is the associated
thermodynamical weight and proportional to the exponential of minus
its scaled (by $\beta$) free energy~\cite{Toki-93,Zhou-2008ctp}.
That is to say, we have now access to the correlations as well as
magnetizations in the form of weighted sum. This weighted sum
actually attenuates the high polarization of the supplied data and a
relatively low inference error is achieved. In fact, SusProp
converges in this case. For both types of state samplings, the
result holds for other random samples, which accounts for the
previously observed large fluctuations of inference error at low
temperatures~\cite{Huang-2010a,Huang-2010b}.

Previous study~\cite{Marinari-2010} emphasized that the inference
error can be drastically reduced by increasing the number of
independent observations, which is consistent with our results in
the sense that state transitions would occur with a higher
probability if the number of sampling time increases. Importantly,
our work discovered further that if the number of sampling time
takes moderate values, the sampling with state transitions can
reduce the inference error while the sampling without state
transitions maintains a high inference error.

\section{Conclusion}
\label{sec_Conc} In conclusion, our study implies that, to lower the
inference error, one should select the most efficient way to sample
the system particularly when the phase space of the original model
develops hierarchically organized structure and the amount of
sampling time is limited (e.g, $2\times 10^{4}$ in our current
simulations). Sampling with state transitions seems to be most
effective to infer the finite-size network structure. In real
neuronal networks, such as retinal network presented with natural
movie stimuli, the coexistence of negative and positive couplings
can lead to frustration and thus the emergence of many metastable
states~\cite{Bialek-09ep,Mora-10ep}. For instance, a recording from
a salamander retina of $40$ neurons showed that several metastable
states appear reproducibly across multiple presentations of the same
movie~\cite{Bialek-09ep}. Our result for the state sampling
dependence of the Hopfield network inference may have some
implications for the inference of real neuronal
networks~\cite{Nature-06,Stevenson-2008,Cocco-09}.


\section*{Acknowledgments}

The author would like to thank Haijun Zhou for support and benefited
from the Kavli Institute of Theoretical Physics China (KITPC)
program "interdisciplinary application of statistical physics and
complex networks". The present work was in part supported by the
National Science Foundation of China (Grant numbers 10774150 and
10834014) and the China 973-Program (Grant number 2007CB935903).



\end{document}